\documentclass{llncs}
\usepackage{graphicx}
\usepackage{pifont}
\usepackage[vlined,ruled]{algorithm2e}
\usepackage[shortlabels]{enumitem}
\usepackage{amssymb}
\usepackage{hhline}
\usepackage{makecell}
\usepackage{xcolor}
\usepackage{amsmath}
\usepackage{microtype}
\usepackage{hyperref}

\usepackage{pgfplots} % pro graf
\usepgfplotslibrary{dateplot} % pro graf
\usepackage{siunitx}

\definecolor{goalcolor}{HTML}{309ee3}
\newcommand{\cmark}{\ding{51}}%
\newcommand{\xmark}{\ding{55}}%
\newcommand{\qmark}{\textbf{?}}
\definecolor{limeGreen}{RGB}{50,205,50}

\pgfplotsset{compat=1.17}

\begin{document}
\title{\texorpdfstring{Implementation of Sprouts: \\a graph drawing game\thanks{T. \v{C}\'{i}\v{z}ek and M. Balko were supported by the grant no. 21/32817S of the Czech Science Foundation (GA\v{C}R).
T. \v{C}\'{i}\v{z}ek was supported by the grant SVV–2020-260 578.
M. Balko acknowledges the support by the Center for Foundations of Modern Computer Science (Charles University project UNCE/SCI/004).
This article is part of a project that has received funding from the European Research Council (ERC) under the European Union's Horizon 2020 research and innovation programme (grant agreement No 810115).
}}{}}
\author{Tom\'{a}\v{s} \v{C}\'{i}\v{z}ek\inst{1}\orcidID{0000-0002-0586-8082} \and Martin Balko\inst{1}\orcidID{0000-0001-9688-9489}}
\authorrunning{T. \v{C}\'{i}\v{z}ek and M. Balko}
\institute{Department of Applied Mathematics,
Faculty of Mathematics and Physics, Charles University, Prague, Czech Republic\\
\email{cizek@kam.mff.cuni.cz, balko@kam.mff.cuni.cz}%\\
%\url{http://www.springer.com/gp/computer-science/lncs}
}
\maketitle              % typeset the header of the contribution
\begin{abstract}
Sprouts is a two-player pencil-and-paper game invented by John Conway and Michael Paterson in 1967.
In the game, the players take turns in joining dots by curves according to simple rules, until one player cannot make a move.
The game of Sprouts is very popular and simple-looking, so it may come as a surprise that there are essentially no AI Sprouts players available.
This lack of computer opponents is caused by the fact that the game hides a surprisingly high combinatorial complexity and implementing it involves fascinating programming challenges.

We overcome all the implementation barriers and create the first user-friendly Sprouts application with a strong artificial intelligence after more than 50 years of the existence of the game.
In particular, we combine results from the theory of nimbers with new methods based on Delaunay triangulations and crossing-preserving force-directed algorithms to develop an AI Sprouts player which plays a perfect game on~up~to~11~spots.
%The abstract should briefly summarize the contents of the paper in 150--250 words.

\end{abstract}

\section{Introduction}

\begin{figure}[!htb]
\centering
\includegraphics[scale=0.65]{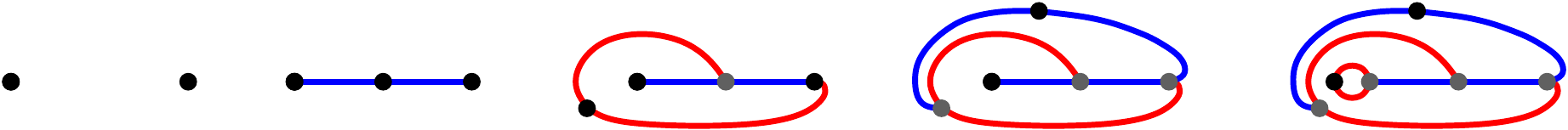}
\caption{An example of a 2-spot game. The first player loses after 4 moves.}
\label{SproutsGameExample}
\end{figure}

\emph{Sprouts} is a 2-player combinatorial paper-and-pencil game with very simple rules.
The game starts with $n$ initial spots and the players alternate in connecting the spots by curves and adding a new spot on each newly drawn curve.
No curve can cross or touch another curve or itself and each spot can be incident to at most three curves.
The first player who cannot make a move loses the game; see an example of a Sprouts game with 2 spots in~Figure~\ref{SproutsGameExample}.

Sprouts was invented by John Conway and Mike Paterson at the University of Cambridge in 1967 with an intention to create a game that is simple to play and yet hard to analyze~\cite{berlekamp,GeniusAtPlay}.
It was later popularized by Gardner~\cite{gardner} in a~\emph{Scientific American} article.
In 1969, Anthony~\cite{Macroscope} mentioned Sprouts in his science-fiction novel \emph{Macroscope} and, few years later, Pritchard~\cite{BestGamesList} listed Sprouts in his compilation of world's best games for two players.
Eventually, Sprouts became very popular with dozens of publications about the game 
and with the \emph{World Game Of Sprouts Association}, which regularly held Sprouts championships. 
Yet, even after more than half a century, there are essentially no implementations of this game with a computer opponent~\cite{InteractiveSprouts}.
This is very surprising at first, given how simple-looking and well-known the game of Sprouts is.
However, the combinatorial complexity of Sprouts is very high and implementing the game thus involves various programming challenges.

We identify three main implementation barriers that cause many difficulties to potential developers.
The first problem is to handle free-form input drawings.
The drawn positions tend to degenerate and become confusing throughout the game and thus one has to come up with a method to maintain them stable and clear.
Second, the number of possible moves can be exponential with respect to the number of initial spots, which makes creating a solid computer opponent very difficult. 
Finally, it is highly nontrivial to synchronize the free-form inputs of a human player with the game representation used by a computer.

In this paper, we overcome all these challenges and create the first user-friendly Sprouts implementation with a strong artificial intelligence.
To do so, we apply techniques from the theory of \emph{nimbers} used by the state-of-the-art Sprouts solver by Lemoine and Viennot~\cite{NimberAnalysis} and combine them with our own \emph{spindle method} for performing computer moves using Delaunay triangulations and crossing-preserving force-directed algorithms.
Our program \emph{Sprouts: A Drawing Game} supports games on up to $20$ spots and contains an AI player that plays perfectly on $n$-spot positions with $n \leq 11$, surpassing all existing Sprouts implementations.
The first version of our program is available at~\cite{program}.

\section{Related work}

In~Table~\ref{ImplementationsOverview}, we list an overview of all implementations of Sprouts that we are aware of.
There are several Sprouts applications that allow to play only against human players. 
This includes \emph{SproutsPlus}~\cite{SproutsPlus}, \emph{Sprouts Game}~\cite{SproutsGame}, and the \emph{University of Utah Sprouts Applet}~\cite{UtahApplet}.
The application \emph{3Graph} by Stefan Reiss~\cite{StefanReiss} is the only Sprouts implementation with AI players.
It plays a perfect game on $n$-spot positions with $n \leq 8$ and supports all features listed in Table~\ref{ImplementationsOverview} except of the remote game.
Although 3Graph is, in our opinion, currently the best Sprouts implementation, there are some places for improvement.
It is not very user-friendly, supports games with only at most 8 spots, and often crashes due to various internal errors.
Browne~\cite{InteractiveSprouts} also mentions \emph{Sprouts - A Game of Maths!}~\cite{Spouts} as a Sprouts application with AI players, but he states that it did not work on any tested device and it is currently unavailable.

\begin{table}[t]
\caption{An overview of the existing Sprouts implementations with the following info: current availability of the program (CA), the support of the free-form inputs (FI), crossing detection (CD), position maintaining (PM), computer opponent (CO), remote game (RG), and the target platform (TP).
Our application \emph{Sprouts: A Drawing Game} supports all the listed features.}
\label{ImplementationsOverview}
\centering
\begin{tabular}{|c|c|c|c|c|c|c|c|}
\hline
& CA
& FI
& CD
& PM
& CO
& RG
& TP\\
\hline
\;{\bfseries Sprouts - A Game of Maths!}\; & \xmark & \qmark & \qmark & \qmark & \cmark & \qmark & iOS \\
\hline
{\bfseries SproutsPlus} & \cmark & \xmark & \xmark & \xmark & \xmark & \cmark & iOS  \\
\hline
{\bfseries Sprouts Game} & \cmark & \cmark & \xmark & \xmark & \xmark & \xmark & iOS \\
\hline
{\bfseries UoU Sprouts Applet} & \cmark & \cmark & \cmark & \xmark & \xmark & \xmark & Applet \\
\hline
{\bfseries 3Graph} & \cmark & \cmark & \cmark & \cmark & \cmark & \xmark & \;Windows\;  \\
\hline
\end{tabular}
\end{table} 

Besides these programs, there are several papers analyzing the implementation challenges of Sprouts~\cite{AirForceRedrawingAlg,InteractiveSprouts,NimberAnalysis} and even some intentions to create Sprouts applications with AI~\cite{UtahApplet,InteractiveSprouts,rocchini}.
For example, Browne~\cite{InteractiveSprouts} mentions creating a complete Sprouts-playing app and investigating AI methods for playing the game at arbitrary sizes.
However, no such result has been published yet.

Although almost no AI players are available, there are some computer Sprouts solvers.
The outcomes of $n$-spot positions for $n \leq 7$ were determined by hand~\cite{BestHandProof,SproutsBet}.
Applegate, Jacobson, and Sleator~\cite{FirstComputerAnalysis} wrote the first computer analysis of Sprouts and successfully determined the winning player of $n$-spot games for each $n \leq 11$.
They also introduced the famous \emph{Sprouts conjecture} which states that each $n$-spot game is winning for the first player if and only if $n$ is equal to 3, 4 or 5 modulo 6.
Lemoine and Viennot~\cite{NimberAnalysis} created the interactive Sprouts position editor \emph{GLOP} that was used to determine all outcomes up to 44 spots and even some outcomes up to 53 spots.
These are the strongest results to date and each of the computed outcomes agrees with the Sprouts conjecture, which, however, still remains open.

\section{Preliminaries}

\begin{figure}[t]
\centering
\includegraphics[scale=1]{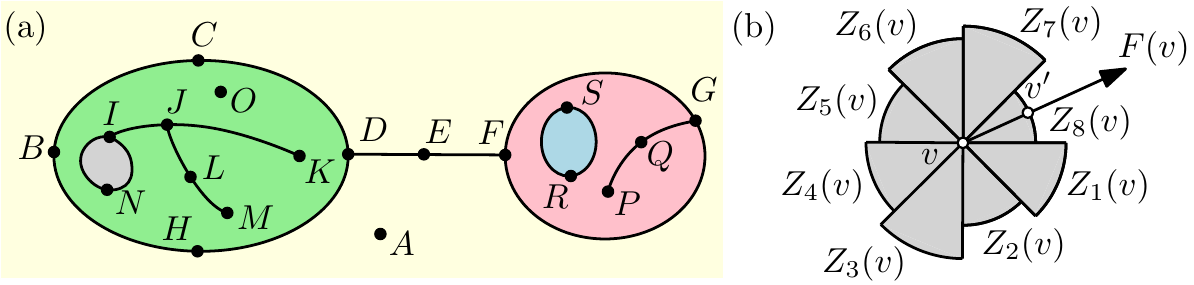}
\caption{(a) A Sprouts position with five regions. (b) A zone $Z(v)$ with a new location $v'$ of the vertex $v$ moved in the direction $F(v)$. }
\label{Zones}
\centering
\end{figure}

We now introduce some notation and show basic properties of Sprouts.
Let $\mathcal{P}$ be a class of plane graphs with maximum degree at most 3, obtained from a finite set of isolated vertices by a sequence of moves that obey the rules of Sprouts.
The plane graphs from $\mathcal{P}$ are called \emph{positions}.
Each vertex of a position corresponds to one of the initial spots or to one of the spots added along the drawn curves.
Edges of a position are formed by the portions of the curves between two spots; see part~(a) of Figure~\ref{Zones}.
For $n \in \mathbb{N}$, the \emph{$n$-spot position} is a position consisting of $n$ isolated vertices.
The number of \emph{lives} of a vertex $v$ in a position~$P$ is $l(v)=3 - {\rm deg}_P(v)$. 
A vertex $v$ is \emph{alive} if $l(v) > 0$ and \emph{dead} otherwise.
The faces of $P$ are called \emph{regions}.
The outer face of~$P$ is called the \emph{outer region} of~$P$ 
and all other faces are called \emph{inner regions} of~$P$.
For a region~$R$ of $P$, we let $P_R$ be the plane subgraph of $P$ induced by vertices that are incident to $R$.
Every connected component of $P_R$ is a \emph{boundary} of~$R$.
For an inner region $R$, the unique boundary incident to the outer region of $P_R$ is the \emph{border boundary} of $R$.
The other boundaries are called \emph{inner boundaries}, including all boundaries of the outer region.
A region $R$ is \emph{alive} if $\sum_{v \in R} l(v) > 1$ and \emph{dead} otherwise.

It is quite easy to show that each game on the $n$-spot ends after at most $3n - 1$ moves and lasts at least $2n$ moves~\cite{berlekamp}.
It follows that Sprouts is a finite game.
In fact, it is an \emph{impartial game}~\cite{OnNumbersAndGames} and thus there exists a winning strategy for one of the players in every position.

\section{Graphical representation}

We assume that the edges of a Sprouts position $P$ are drawn as piece-wise linear arcs and that $P$ is contained in $[0,1]^2$, which represents the playing area.
Vertices of the \emph{graphical representation} $gr(P)$ are the endpoints of the line segments forming the piece-wise linear arcs representing the edges of $P$ or Sprouts spots that stand alone somewhere in a region of $P$.
The vertices in the latter case are called \emph{singletons}.
The \emph{game vertices} correspond to the real Sprouts spots.
They are either singletons or the endpoints of the edges of $P$.

In $gr(P)$, we represent an edge $e$ of $P$ by a sequence $gr(e)$ of vertices of $e$ starting and ending with the game vertices that are the endpoints of $e$.
The ordering of $gr(e)$ is determined by the order in which $e$ was drawn.
The line segment connecting two consecutive vertices of $gr(e)$ is called a \emph{small edge}.
A boundary $\beta$ of $P$ is represented by a circular list of the sequences representing the edges of $\beta$.
It is important for the move insertion that all inner boundaries in a region of $P$ are oriented in the same way while the orientation of the border boundary is opposite.
A region $R$ is represented by a set of the lists representing the boundaries of $P$ organized into a tree-like hierarchical structure, which is used to identify the region that contains a given point.

The graphical representation is used for correct insertion of the moves, crossing detection, and for redrawing of positions. 
Due to space limitations, we describe only the redrawing algorithm in detail.

\subsection{Redrawing algorithm}
\label{subsec-redrawing}

To keep the freely-drawn positions stable and clear, we implemented a modification of the redrawing algorithm \emph{PrEd} by Bertauls~\cite{PrEd} and its improved version \emph{ImPrEd} by Simonetto et al.~\cite{ImPrEd}.
PrEd and ImPrEd are iterative force-directed algorithms that improve a given graph drawing while preserving its edge-crossing properties.
At each iteration, we compute a force $F(v) \in \mathbb{R}^2$ for each vertex $v$ of a position $P$ and then move $v$ in the direction $F(v)$.
The amplitude of each move is restricted so that the edge-crossing properties are preserved.

We let $V$ be the vertex set of $P$, 
$E$ be the set of edges of $P$, and $E_s$ be the set of small edges of~$P$.
We define the set $S$ of four additional static edges around the game board $[0, 1]^2$, which prevent vertices to move outside of the playing~area.

We use three different forces between vertices and small edges: the attraction force $F^a$ between vertices that are connected by a small edge, the repulsion force $F^r$ between pairs of vertices, and the repulsion force $F^e$ between vertices and small edges.
To achieve better results, we count the force $F^r$ only between vertices that do not lie on the same edge of $P$ unless they are both game vertices.
We use three different constants $\beta_{u, v}$, $\gamma$, and $\delta$ that customize the balance of these forces.
The constant $\gamma$ is the optimal distance of edges and $\delta$ is equal to the optimal length $l_{opt}$ of a small edge.
The value of $\beta_{u, v}$, which depends on two vertices $u$ and $v$, is mainly used to strengthen the repulsion force between $u$ and $v$ if $u$ and $v$ are adjacent game vertices.
The precise values of the attraction force $F^a\colon \binom{V}{2} \to \mathbb{R}^2$ and the repulsion force $F^r\colon \binom{V}{2} \to \mathbb{R}^2$ are
\[
F^a(u, v) = \frac{\|u - v\|}{\delta}(u - v) \;\;\; \text{ and } \;\;\;\; F^r(u, v) = \left( \frac{\beta_{u, v}}{\|u - v\|} \right)^2 (v - u),
\]
where $F^r(u, v) = (0,0)$ if $\|u-v\| \geq\beta_{u,v}$. The force $F^e\colon V \times E_s \to \mathbb{R}^2$ is given by
\[
F^e(v, ab) = \left\{
\begin{array}{ll}
      \frac{(\gamma - \|v - v_{ab}\|)^2}{\|v - v_{ab}\|}(v - v_{ab}) & v_{ab} \in ab, \|v - v_{ab}\| < \gamma \\
      (0,0) & \text{otherwise,}\\
\end{array}
\right.
\]
where $v$ is a vertex disjoint from the edge $ab$ and the point $v_{ab}$ is the projection of the vertex $v$ onto the line determined by the small edge $ab$.
The overall force $F\colon V \to \mathbb{R}^2$ is computed for each vertex $v$ as follows
\[
F(v) = \sum_{\substack{u \in V\\ uv \in E_s}} F^a(u, v) + \sum_{\substack{u \in V \setminus \{v\}\\ \nexists e \in E \colon uv \in e}} F^r(u, v) + \sum_{\substack{ab \in E_s \cup S\\v \notin ab}} F^e(v, ab) - \sum_{\substack{u,w \in V\\ vw \in E_s\\ u \notin vw}} F^e(u, vw).
\]

After each force is computed, the vertices are moved in the direction $F(v)$.
To preserve edge-crossings, we restrict the amplitude of these movements in the same way as in PrEd~\cite{PrEd}.
For each vertex~$v$, we define a \emph{zone} $Z(v)$ as a union of eight circular sectors $Z_1(v), \dots, Z_8(v)$ with radii $r_1(v), \dots, r_8(v)$, respectively; see part~(b) of Figure~\ref{Zones}.
At the end of each iteration, we compute the zone of each vertex $v$, find a sector $Z_i(v)$ that intersects the computed force $F(v)$, and then we move $v$ in the direction $F(v)$ by $\min \{\|F(v)\|, r_i(v)\}$.
We set the radii $r_1(v), \dots, r_8(v)$ so that the edge-crossings are preserved.
To compute the radii of a zone $Z(v)$, we only consider small edges $ab$ of $P$ that are disjoint from $v$.
Based on each such a small edge $ab$, we restrict the appropriate radii of the vertices $v$, $a$, and $b$.
The initial values of the radii are set to a maximum size $M_{max}$ of a movement within a single iteration.
Updating the radii is then done as in~\cite{PrEd}, which implies that the same proof of the correctness still applies here.

To avoid having too short or too long edges, we merge too short small edges together after each iteration.
If there are two adjacent small edges $uv$ and $vw$ with the distance $\|u - w\|$ below a certain value $l_{mer}$, then the small edges $uv$ and $vw$ are replaced by a new small edge $uw$.
To prevent crossings, we merge $uv$ and $vw$ only if there is no vertex of $P$ in the triangle $uvw$.
We also subdivide the small edges of $e$ whose length is greater than a certain value $l_{sub}$.

To speed up the computation, we update zones of vertices only for small edges that lie in the same region.
Moreover, we use a graphical data structure \emph{quadtree} as proposed by Simonetto et al.~\cite{ImPrEd}.
The use of quadtrees leads to a sublinear time of searching for vertices that are close to a given point.
Since the amplitude of the movements of vertices is bounded by the finite value $M_{max}$, we update each zone $Z(v)$ using only small edges $s$ whose distance from $v$ is less than $2 M_{max}$ using quadtrees.
We also compute forces only between elements that are close enough.
More precisely, we compute the repulsion force $F^r(u, v)$ between vertices $u$ and $v$ only if their distance is less than $\beta_{u, v}$.
Similarly, we compute the edge repulsion force $F^e(v, ab)$ only if the distance between the vertex $v$ and the small edge $ab$ is less than $\gamma$.
To determine close pairs of two vertices and of a vertex and a small edge, we implemented quadtrees for vertices and for edges.

\begin{figure}[t]
\centering
\begin{tabular}{ccc}
\includegraphics[scale=0.15]{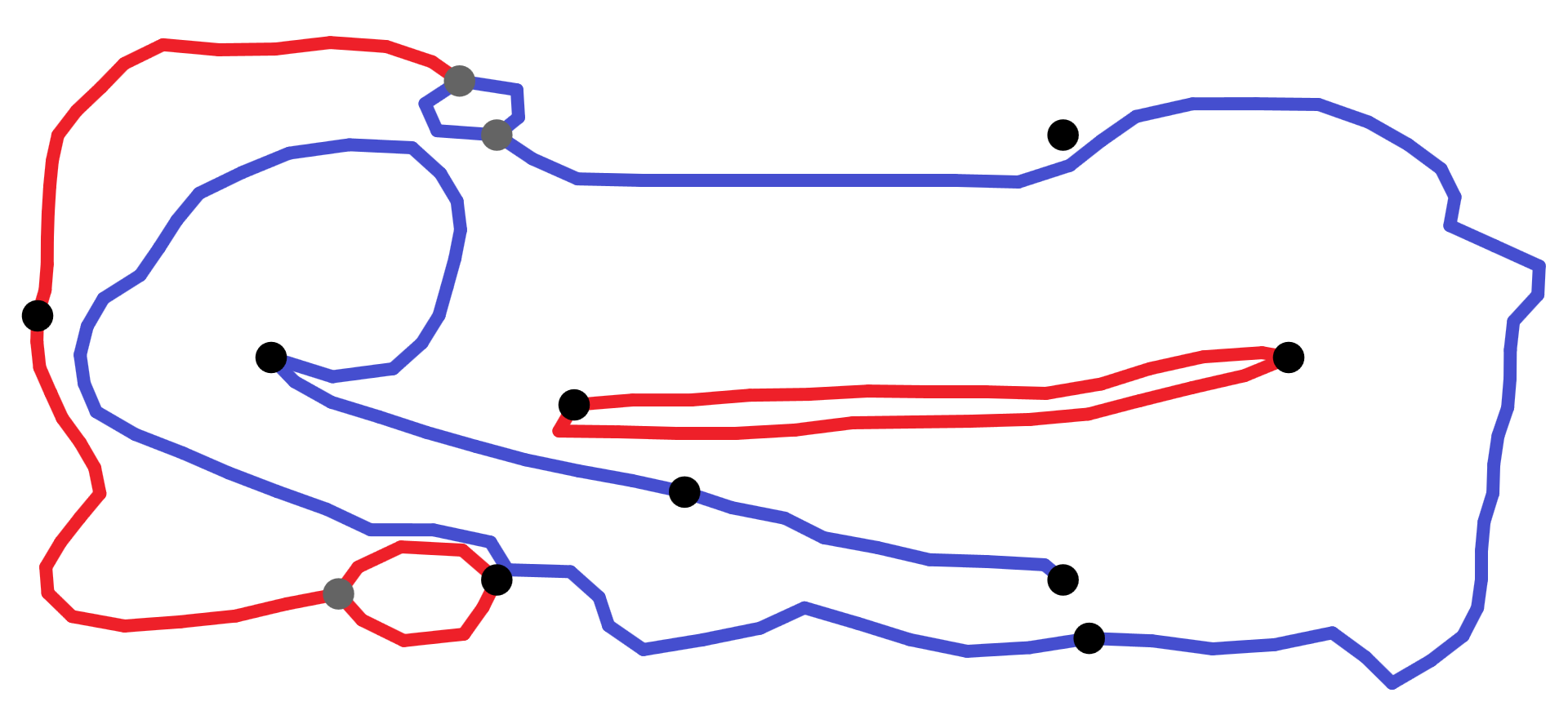} &
    &
    \includegraphics[scale=0.14]{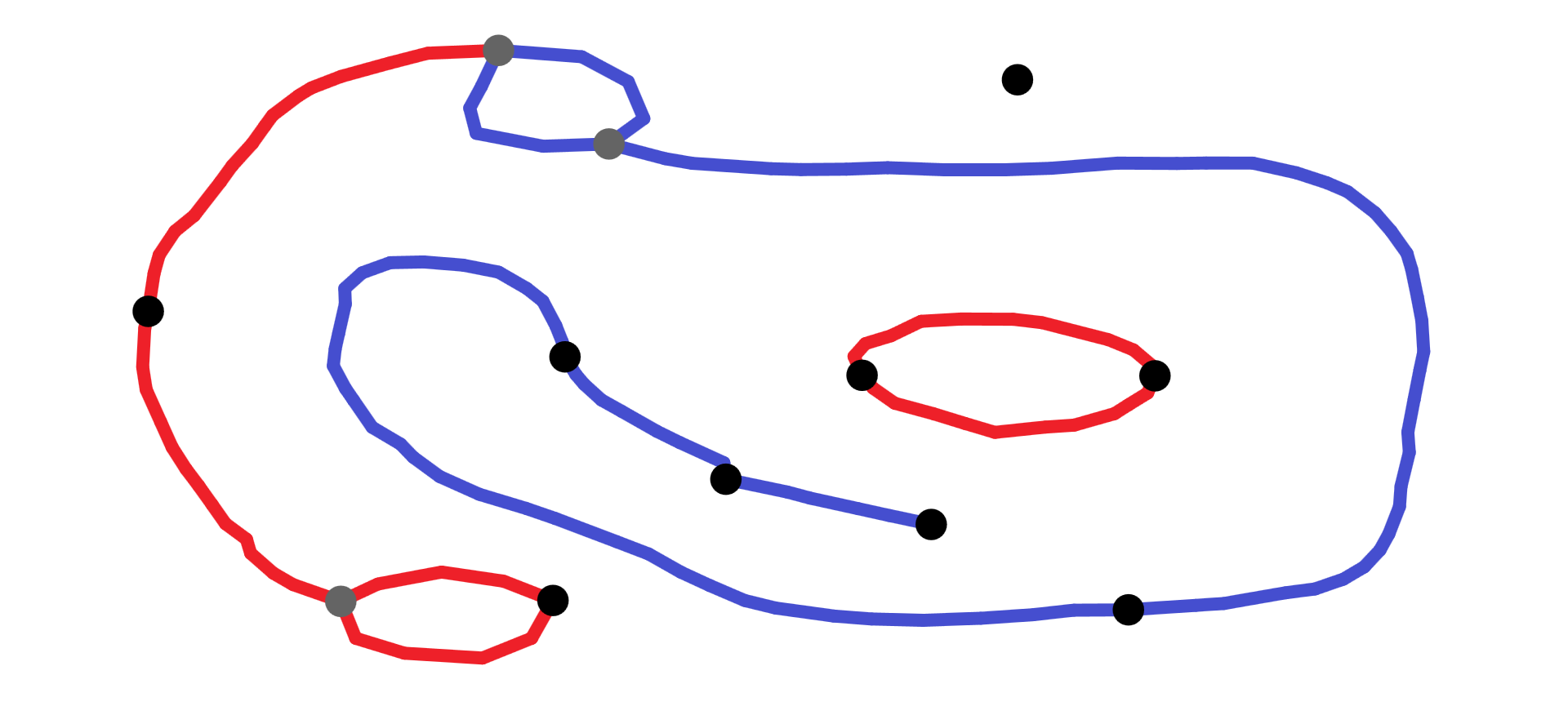} \\
    \normalsize{Before redrawing.}  && \normalsize{After redrawing.} \\
\end{tabular}
\caption{Redrawing a Sprouts position with 6 initial spots using 30 iterations.}
\label{RedrawingExample}
\end{figure}

Our redrawing algorithm is suitable for real-time computations in games with up to 20 spots; see Figure~\ref{RedrawingExample}.
We use 30 iterations with a time limit after which the redrawing is terminated even if not all the iterations have finished.
We also draw all curves with \emph{Bezier splines} to make the moves look smoother.

\section{Computer opponent}

To implement a strong computer opponent, we apply methods used in the currently best Sprouts solver \emph{GLOP}~\cite{NimberAnalysis}.
First, we encode each Sprouts position~$P$ in a compact way using a \emph{string representation} $sr(P)$, which was considered by many authors~\cite{FirstComputerAnalysis,InteractiveSprouts,NimberAnalysis}. 
We describe the string representation and the methods very briefly, as we use them in the same way as Lemoine and Viennot~\cite{NimberAnalysis}.

The string representation $sr(P)$ of a position $P$ is a string of game vertices, which are denoted by capital letters.
Each boundary $\beta$ forms a block of consecutive letters in $sr(P)$ in the order we meet the corresponding game vertices when traversing $\beta$.
The boundaries are separated by dots in $sr(P)$. 
The regions are represented by consecutive blocks of boundaries  separated by $|$ in $sr(P)$.
The position $P$ is also split into \emph{lands} which are mutually independent parts of $P$.
The lands are separated by $+$ in $sr(P)$.
The strings of inner boundaries within the same region have the same orientation while the orientation of the string of the border boundary is opposite.
For example, the position $P$ from Figure~\ref{simplification} can be represented by the string \texttt{A.BCDEFGFEDH|DCBH.IJKJLMLJIN.O|IN+PQGFGQ.RS|RS}.

\subsection{String simplification}
\label{subsec-simplification}

A single Sprouts position can be represented by many different strings which leads to repetitive computations.
Therefore, we simplify the string representation by applying two methods called the \emph{string reduction} and the \emph{string canonization}.

\begin{figure}[htb]
\centering
\includegraphics[scale=1]{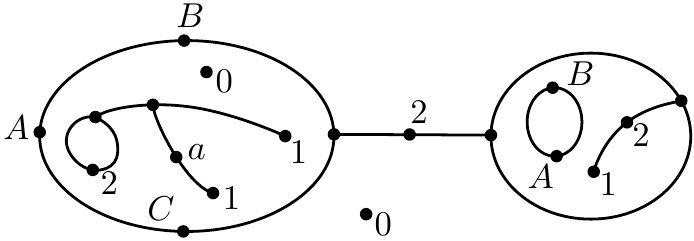}
\caption{The position from part~(a) of Figure~\ref{Zones} after the string reduction, which produces the string \texttt{0.AB2C|BAC.1a1a2.0+12.AB|AB}.
}
\label{simplification}
\centering
\end{figure}

\paragraph{String reduction}
The string reduction simplifies strings using five steps.
After the reduction, the string \texttt{A.BCDEFGFEDH|DCBH.IJKJLMLJIN.O|IN|PQGFGQ.RS|RS} of the position from part~(a) of Figure~\ref{Zones}, which is not partitioned into lands, becomes \texttt{0.AB2C|BAC.1a1a2.0+12.AB|AB}; see Figure~\ref{simplification}.

\begin{enumerate}
    \item \textbf{Delete dead parts}: We delete all dead vertices, then all boundaries that newly become empty, and then all dead regions and all empty lands.
    
    \item \textbf{Apply generic names}: We rename each singleton to \texttt{0} and each vertex with two lives to \texttt{1}.
    After merging vertices that occur twice in a row, we rename each letter that occurs only once in the string to \texttt{2}.
    
    \item \textbf{Split lands}: We split the position into lands.
    
    \item \label{RenameLettersStep} \textbf{Rename letters}: 
    We rename vertices contained in a single boundary starting from \texttt{a} in the order they occur in the string.
    Then we rename all other letter vertices within a single land starting from \texttt{A} in the order they occur.
    
    \item \textbf{Merge boundaries}: For every region $R$ with $\sum_{v \in R}l(v) \leq 3$, we merge all boundaries of $R$. This considerably reduces the number of strings.
\end{enumerate}

\paragraph{String canonization}
Even after the string $sr(P)$ is reduced, there can be many reduced strings representing $P$ since they might differ only by reordering of the boundaries, regions, lands and by relabeling of the vertices.
A \emph{canonization} selects the lexicographically minimal string representing $P$ and thus makes the representation unique.
However, the canonization takes too much running time.
We thus perform only the \emph{pseudocanonization} used by Lemoine and Viennot~\cite{NimberAnalysis}.
This does not make the string unique, but it considerably reduces the number of strings.
For example, the string \texttt{0.AB2C|BAC.1a1a2.0+12.AB|AB} then becomes \texttt{0.12a1a.ABC|0.2ABC+12.AB|AB}, which we would get in the full canonization.

\subsection{Computation of outcomes}
\label{subsec-outcomes}

Since Sprouts is an impartial game, we can determine the outcome of each position $P$.
If the first player has a winning strategy, then $P$ is \emph{winning} and its outcome is $Win$. 
Otherwise, $P$ is \emph{losing} with the outcome $Loss$.
To determine the outcome of $P$, we apply the theory of nimbers~\cite{berlekamp}.
The \emph{nimber} $|P|$ of a position $P$ is the smallest non-negative integer that is not a nimber of a child of $P$.
It follows that $P$ is losing if and only if $|P|=0$.
The nimbers speed up the computation as positions tend to consist of several lands and the outcome of $P$ can be obtained by merging the nimbers of the lands of $P$ according to the following result.

\begin{theorem}[\cite{berlekamp,NimberAnalysis}]
\label{tmh-nimber}
The nimber of a Sprouts position $P$ consisting of lands $L_1, \dots,  L_n$ is $|P| = |L_1| \oplus \cdots \oplus |L_n|$, where $\oplus$ denotes the bitwise exclusive or.
\end{theorem}

Instead of computing the outcome of $P$, we compute the outcome of a \emph{couple} $(P+n)$ which consists of a parallel game of Sprouts on $P$ and the game of Nim~\cite{berlekamp} on a single heap of $n$ objects.
This corresponds to computing nimbers of Sprouts positions which allows us to apply Theorem~\ref{tmh-nimber} and split positions into lands on which we proceed independently.
We apply Algorithm~\ref{ComputingNimber} to efficiently compute the nimber of $P$ using that the outcome of $(P+n)$ is Loss if and only if $|P| = n$.

\SetKwProg{Fn}{Function}{}{}
\begin{algorithm}[t]
\Fn{ComputeNimber($P$)}{
    $n \leftarrow 0$\;
    \While{$True$}
    {
        \If{$ComputeOutcome(P+n)$ is $Loss$}
        {
            \Return $n$\;
        }
        
        $n \leftarrow n + 1$\;
    }
}
\caption{Computing the nimber of a position $P$}
\label{ComputingNimber}
\end{algorithm}

We apply the Alpha-beta pruning algorithm to compute the outcome of $(P+n)$; see Algorithm~\ref{ComputingOutcomeWithNimbers}.
The set of children of $(P+n)$ is equal to the union of $\{\,(C, n) \colon C\text{ is child of }P\,\}$ and $\{\,(P, m) \colon m < n \,\}$.
Note that the outcome of $P$ is the same as the outcome of $(P+0)$.
We only store losing couples, which means that we only store positions whose nimber is known.
Also, we store only single lands in our database as we are computing the lands separately.

\SetKwProg{Fn}{Function}{}{}
\begin{algorithm}[t]
\Fn{ComputeOutcome($P+n$)}{
    merge nimbers of lands whose nimbers are stored in the database with the nimber part of $(P+n)$ using the bitwise exclusive or $\oplus$\;
    compute the unknown nimbers of all lands remaining in $P$ except of one  land $P'$ and merge them with the nimber part of $(P+n)$\; 
    \BlankLine
    $(P'+n') \leftarrow$ the updated couple $(P+n)$ after the previous steps\;
    \ForEach{child $C$ of $(P'+n')$}
    {
        $outcome \leftarrow ComputeOutcome(C)$\;
        \If{$outcome$ is $Loss$}
        {
            \Return $Win$\;
        }
    }
    
    store $(P'+n')$ into the database\;
    \Return $Loss$\;
}
\caption{Computing the outcome of a couple $(P+n)$}
\label{ComputingOutcomeWithNimbers}
\end{algorithm}

\subsection{Creating the computer opponent}
\label{subsec-AI}

The computer opponent tries to compute the outcome of a position $P$.
If it is Win, then he plays a move that leads to a losing child.
Otherwise the outcome is Loss and all children are winning. The AI player can then select an arbitrary move.
To make the exploration of the game tree faster, we use known techniques such as a suitable \emph{children priorities}~\cite{NimberAnalysis} and \emph{boundary matching} on singletons~\cite{InteractiveSprouts}.

We also implemented new features to improve the AI player.
To make sure that the perfect AI never takes too long in finding a best move, we support \emph{databases of pre-trained positions}.
Unlike Sprouts solvers, the AI player has to consider all moves in the human player's turns so it is necessary to explore the game tree in a much larger width.
This is a problem, as the game trees are very large.
For example, already the tree of the 6-spot position contains 393103 strings.
We solve this by restricting the set of possible moves of the AI player so that we do not have to explore the whole tree.
The computer also selects children that are simpler to analyze in wining positions and children with proportionally most losing children in losing positions to increase a chance of opponent's mistake.

Lemoine and Viennot~\cite{NimberAnalysis} also suggested to implement \emph{distributed computing} for determining the outcomes.
We implemented this improvement with threads using a global database that is equipped by synchronization primitives to prevent deadlocks.
The outcome computation can be much faster with more threads.

\section{Drawing a computer move}

Here, we synchronize the graphical representation $gr(P)$ with the string representation $sr(P)$ of a position $P$ so that we can draw a computer move found with $sr(P)$ into $gr(P)$.
This is one of the most difficult steps we had to deal with and as far as we know, it is not fully described in the literature.
Browne~\cite{InteractiveSprouts} sketched out the idea of using \emph{Delaunay triangulations} and \emph{Voronoi diagrams}.
Although his solution works for the $n$-spot positions, there are several missing parts for more complicated positions.
So we apply our own new \emph{spindle method}.

\begin{figure}[!htb]
\centering
\includegraphics[scale=0.7]{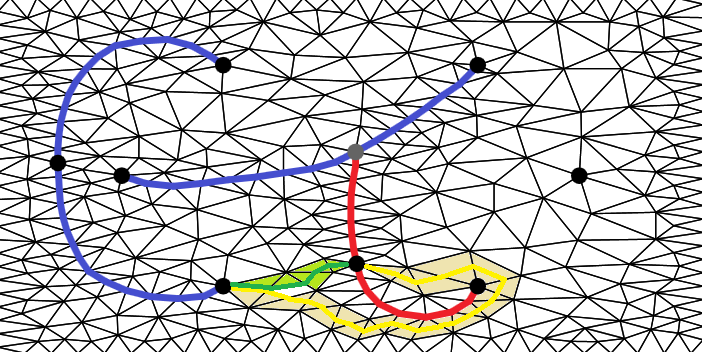}
\caption{A region triangulated by a CCDT $\mathcal{T}$ and of two paths (surrounded by shaded triangles) in $G_{\mathcal{T}}$ that connect different vertex occurrences of the same vertices.}
\label{SearchingPathExample}
\end{figure}

We also use Delaunay triangulations for computer's drawing.
The \emph{constrained conforming Delaunay triangulation} (CCDT) of a region of $P$ is a constrained Delaunay triangulation using \emph{Steiner points} to meet given constraints on the minimum angle and the maximum area of the triangles.
We use the CCDT to triangulate a region whose edges are all constrained; see Figure~\ref{SearchingPathExample}.
Let $p_1$ and $p_2$ be two non-Steiner points of a CCDT~$\mathcal{T}$.
We define a plane graph $G_\mathcal{T}=G_\mathcal{T}(p_1,p_2)$ by letting the vertices of $G_\mathcal{T}$ be the points $p_1$, $p_2$ and the midpoints of all non-constrained edges of $\mathcal{T}$.
Two vertices are connected by an edge in $G_\mathcal{T}$ if they lie in the same triangle of $\mathcal{T}$ but not on the same edge of $\mathcal{T}$.
If $p_1$ and $p_2$ lie in the same triangle $T$, then we connect them through the center of gravity of $T$ in $G_\mathcal{T}$.

We distinguish two types of moves depending on whether they connect vertices from different boundaries or from the same boundary of $P$.
Consider two different boundaries $A_1 \cdots A_i \cdots A_m$ and $B_1 \cdots B_j \cdots B_n$ of $P$.
The \emph{double-boundary move} that connects two \emph{vertex occurrences} $A_i$ and $B_j$ creates a new boundary $A_1 \cdots A_i Z \allowbreak B_j \cdots B_n \allowbreak B_1 \cdots B_j Z A_i \cdots A_m$ where $Z$ is the newly added vertex.
Let $A_1 \cdots A_i \cdots \allowbreak A_j \cdots A_n$ be a boundary in a region $R$ with boundaries partitioned into sets $\mathcal{B}_{major}$, $\mathcal{B}_{minor}$, and $\{\beta\}$.
The \emph{single-boundary move} that connects (not necessarily different) vertex occurrences $A_i$ and $A_j$ and separates the boundaries $\mathcal{B}_{major}$ from the boundaries of $\mathcal{B}_{minor}$ splits $R$ into the \emph{major region} $A_i \cdots A_j Z.\mathcal{B}_{major}$ and into the \emph{minor region} $A_1 \cdots A_i Z A_j \cdots A_n.\mathcal{B}_{minor}$.

\paragraph{Drawing a double-boundary move}
To draw a double-boundary move $m$ between two vertex occurrences $A_i$ and $B_j$ in a region $R$, we construct a triangulation $\mathcal{T}$ of $R$ and we let $m$ be the shortest path in $G_\mathcal{T}(A_i,B_j)$ between $A_i$ and $B_j$.

\begin{figure}[!htb]
\centering
\begin{tabular}{ccc}
\includegraphics[scale=1]{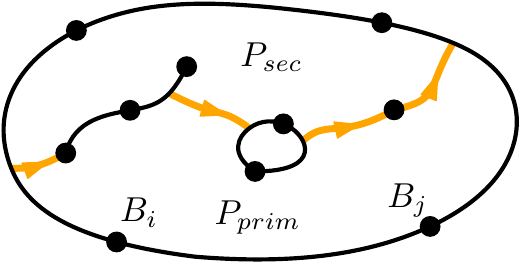} & \;\;\;\;\;\;
    &
    \includegraphics[scale=1]{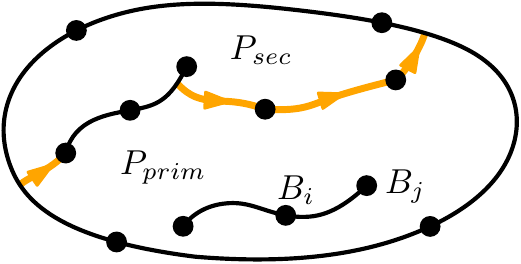} \\
    A border boundary move. && An inner boundary move. \\
\end{tabular}
\caption{Spindles (heavier orange curves) that split the region into $P_{prim}$ and~$P_{sec}$.}
\label{Spindle}
\end{figure}

\begin{figure}[!htb]
\centering
\begin{tabular}{ccc}
\includegraphics[scale=1]{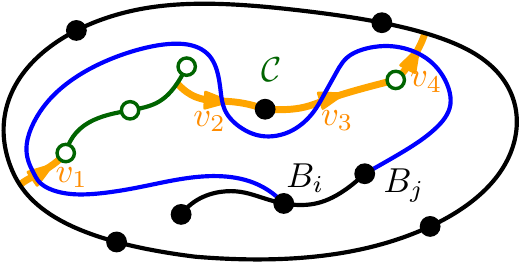} & \;\;\;\;\;\;
    &
    \includegraphics[scale=1]{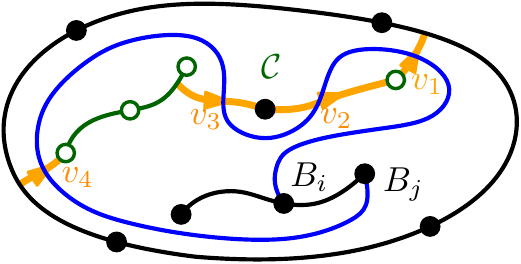} \\
    The enfolding of $\mathcal{C}$. && The reversed enfolding of $\mathcal{C}$. \\
\end{tabular}
\caption{The enfoldings of the boundaries from $\mathcal{C}$ (green curves with empty discs). }
\label{Enfolding}
\end{figure}

\paragraph{Drawing a single-boundary move}
Consider a single-boundary move $m$ connecting vertex occurrences $B_i$ and $B_j$ with $i \leq j$ on a boundary $\beta$ of a region $R$ with boundaries $\mathcal{B}$ that splits the boundaries $\mathcal{B} \setminus \{\beta\}$ into a major partition $\mathcal{B}_{major}$ and a minor partition $\mathcal{B}_{minor}$. Drawing of $m$ is much more complicated since we have to correctly split $\mathcal{B}$ into $\mathcal{B}_{major}$ and $\mathcal{B}_{minor}$.
The first step is to connect all the inner boundaries from $\mathcal{B} \setminus \{\beta\}$ by a curve called \emph{spindle} that starts and ends in the border boundary (or the border of the playing area if $R$ is the outer region); see Figure~\ref{Spindle}.
Then we intertwine $m$ with the spindle so that the partitions $\mathcal{B}_{major}$ and $\mathcal{B}_{minor}$ are on the correct sides of $m$; see Figure~\ref{Enfolding}.
Intertwining $m$ also uses triangulations and requires a lot of technical steps that are sketched below.

\paragraph{Setting up the spindle}
The spindle starts at an arbitrary vertex occurrence of the border boundary and leads to the closest vertex of an inner boundary from $\mathcal{B} \setminus \{\beta\}$.
Then it continues from a vertex of the last visited boundary to the closest vertex of a non-visited inner boundary from $\mathcal{B} \setminus \{\beta\}$ until we visit all the inner boundaries from $\mathcal{B} \setminus \{\beta\}$.
We end the spindle by connecting it to the closest non-visited vertex occurrence of the border boundary; see Figure~\ref{Spindle}.
The spindle divides the border polygon of $R$ into the \emph{primary polygon} $P_{prim}$, which is the polygon that contains $\beta$ or the occurrences $B_i$ and $B_j$ if $\beta$ is the border boundary, and the \emph{secondary polygon} $P_{sec}$.
The orientation of the spindle is opposite to the orientation induced by the counterclockwise orientation of $P_{prim}$.

\paragraph{Intertwining the spindle}
Let $\mathcal{C}$ be a set of some of the inner boundaries visited by the spindle $s$. 
We intertwine the move $m$ with $s$ using an \emph{enfolding} of $\mathcal{C}$; see Figure~\ref{Enfolding}.
To enfold $\mathcal{C}$, we first draw a curve in $P_{prim}$ from $v_0 = B_i$ to a vertex $v_1$ of the first segment of $s$ that precedes a boundary from $\mathcal{C}$ with respect to the orientation of~$s$.
In $P_{sec}$, we then connect $v_1$ with a vertex $v_2$ of the first segment of $s$ that precedes a boundary not in $\mathcal{C}$.
We continue connecting $v_i$ with $v_{i+1}$ like this alternatingly in $P_{prim}$ and $P_{sec}$ until we get at the end of $s$.
At the end, if we should continue drawing in $P_{sec}$, we just draw a curve to the last segment of $s$ to get back to $P_{prim}$.
Finally, we draw a curve in $P_{prim}$ from the last connected vertex to $B_j$.
If $\mathcal{C}$ is empty, we connect $B_i$ with the first vertex of the last segment of $s$ in $P_{prim}$, then we go to the last vertex of the last segment of $s$ in $P_{sec}$, and then we return to $B_j$ in $P_{prim}$.
In a \emph{reversed enfolding} of $\mathcal{C}$, we intertwine $s$ in the opposite direction.

\begin{figure}[!hb]
\centering
\begin{tabular}{ccc}
    \includegraphics[scale=1]{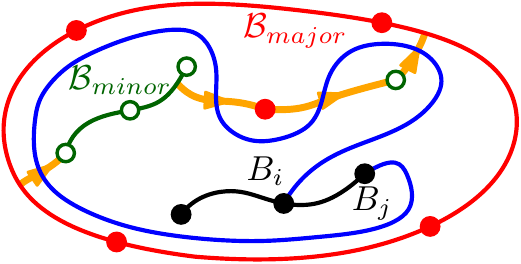} & \;\;\;\;\;\;
    &
    \includegraphics[scale=1]{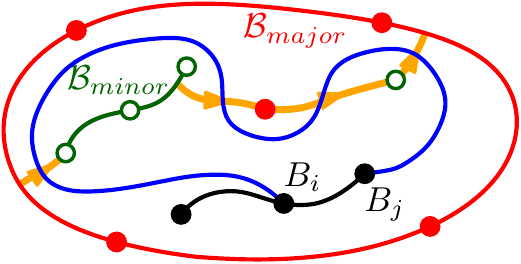}\\
    (a) The reversed enfolding of $\mathcal{B}_{minor}$. & \;\;\;\;\;\;
    & (b) The enfolding of $\mathcal{B}_{minor}$. \\[10pt]
    \includegraphics[scale=1]{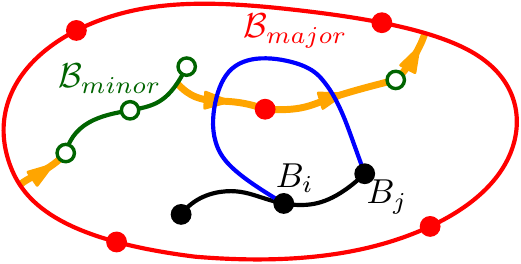}  & \;\;\;\;\;\;
    & \includegraphics[scale=1]{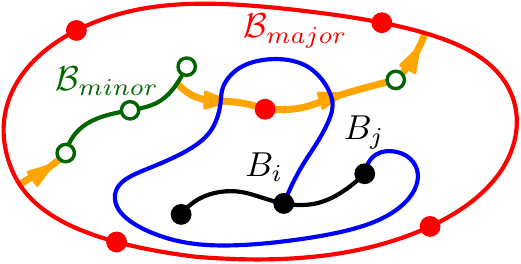}
    \\
     (c) The enfolding of $\mathcal{B}_{major}$. & \;\;\;\;\;\;
    & (d) The reversed enfolding of $\mathcal{B}_{major}$.\\
\end{tabular}
\caption{The (a) correctly and (b--d) incorrectly chosen enfoldings for drawing a move (blue) from $B_i$ to $B_j$ with the partitions $\mathcal{B}_{major}$ (full red) and $\mathcal{B}_{minor}$ (empty green). }
\label{IncorrectEnfolding}
\end{figure}

\paragraph{Choosing the right enfolding}
We have four options how to enfold the partitions.
We can choose $\mathcal{C}$ as the set of inner boundaries from $\mathcal{B}_{major}$ or from $\mathcal{B}_{minor}$.
We can also apply either the enfolding or the reversed enfolding.
However, since the enfolded boundaries $\mathcal{C}$ always lie in the major region of $m$ whereas the reversely enfolded boundaries $\mathcal{C}$ always lie in the minor region and since the border boundary cannot be enfolded nor reversely enfolded, we are left with a single option for enfolding.
We enfold $\mathcal{C} = \mathcal{B}_{major}$ if $\mathcal{B}_{major}$ does not contain the border boundary and we reverse enfold $\mathcal{C} = \mathcal{B}_{minor}$ otherwise; see Figure~\ref{IncorrectEnfolding}.

\paragraph{Optimizing moves}
We use various techniques to make the moves nicer, as they should resemble moves drawn by a human player.
For example, we take shorter equivalent moves if the border boundary is dead, we enfold only sets of close singletons as singletons are interchangeable, and we handle empty single-boundary moves  separately. 
Finally, we note that some technical steps in the analysis of intertwining (for example the case $B_i=B_j$) are omitted in this extended abstract.

\section{Conclusions and future work}

\begin{figure}[!htb]
\centering
    \includegraphics[scale=1]{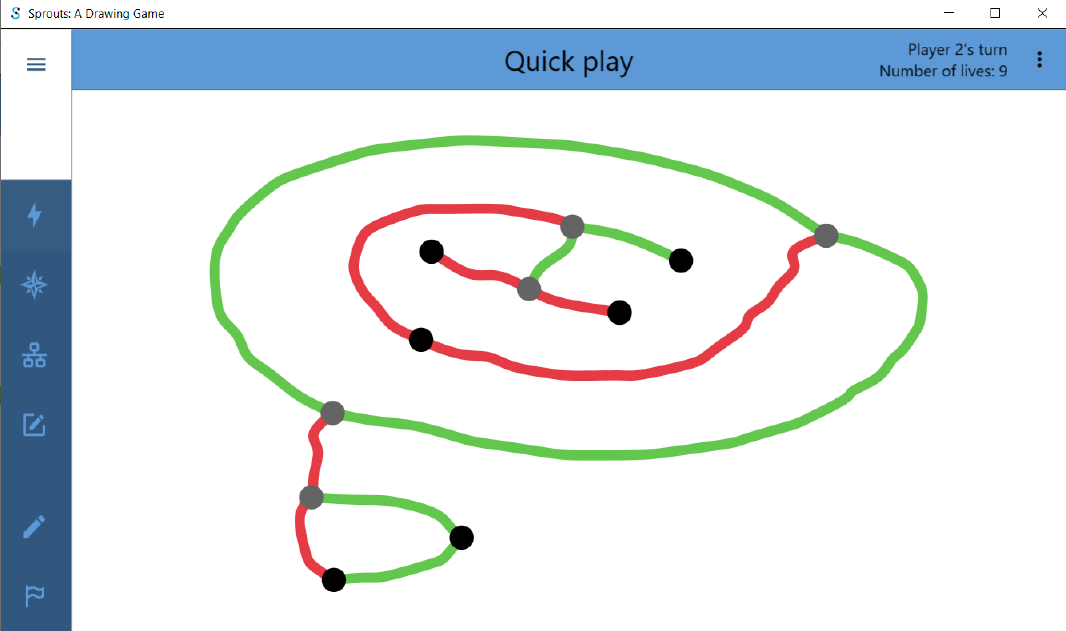} 
\caption{A screenshot from our application \emph{Sprouts: A Drawing Game}.}
\label{SproutsApplication}
\end{figure}

We applied all the techniques described above to implement the Windows application \emph{Sprouts: A Drawing Game} \cite{program} that allows to play against a computer opponent; see Figure~\ref{SproutsApplication} for a screenshot.
Our program supports various forms of the game.
We have a \emph{Campaign mode} consisting of 150 positions derived from the $n$-spot positions with $n \leq 11$ where one can play against a perfectly playing computer opponent or against three other levels of AI.
In a \emph{Quick play mode}, it is possible to play against various computer opponents or against a human opponent on the $n$-spot positions with $n \leq 20$ or on custom positions made in the \emph{Custom maps mode}.
The program is not restricted to only local games but it also supports a \emph{Remote game mode} where the players can connect to a Sprouts server and play against each other remotely.  

\begin{figure}[!htb]
\centering
\begin{tikzpicture}
\begin{axis}[
	width=12.8cm,
	height=8cm,
	legend style={at={(0.02,0.96)},anchor=north west},
	xmin=-0,
    xmax=12,
	xtick=data,
	xticklabels={(7,4), (7,6), (8,2), (8,4), (8,6), (9,2), (9,4), (9,6), (10,2), (10,4), (10,6)},
	ymin= -3000,
    ymax= 56000],
	
\addplot[color=orange!30!red,mark=*] table {GlopData.dat};
\addplot[color=green!80!black,mark=triangle*] table {1ThreadData.dat};
\addplot[color=blue,mark=square*] table {4ThreadData.dat};

\legend{GLOP, SADG (1 thread), SADG (4 threads)}

\end{axis}
\end{tikzpicture}
\caption{Comparison of the computation times of \emph{GLOP} and \emph{Sprouts: A Drawing Game} (SADG). 
Every column $(i,j)$ contains the average times (in milliseconds) of determining the outcome of 3 games starting on the $i$-spot position with $j$ randomly selected moves with empty databases of pre-trained moves.
All the computations were performed on a computer with the Intel(R) Core(TM) i7-6700HQ CPU running at 2.60GHz with 16GB of RAM.}
\label{SproutsApplicationPerformance}
\end{figure}
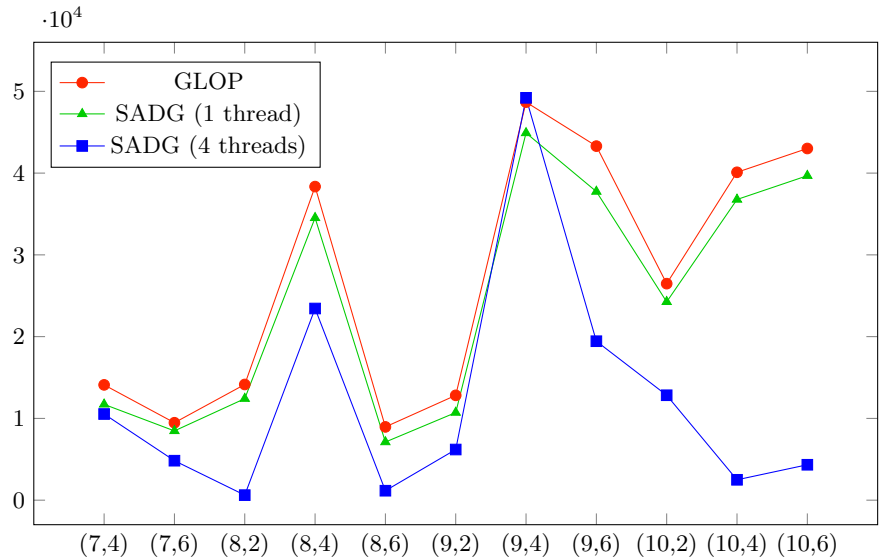

All the computed outputs agree with the results of \emph{GLOP}.
In particular, we obtained the same sizes of the game trees of the $n$-spot positions with $n \leq 6$.
Moreover, the computation of outcomes with our program is slightly faster than with \emph{GLOP}, even when using a single thread; see Figure~\ref{SproutsApplicationPerformance}.
With four threads, the computation can speed up significantly.
We encode positions using the same notation as \emph{GLOP} so its databases can be used in our program as well.

We plan to further improve our application, for example, we would like optimize the enfolding of singletons to make the computer moves even more natural.
Another possible plan is to employ the distributed computations of outcomes in an implementation of a new Sprouts solver in order to determine the outcomes of new $n$-spot positions and to tackle the Sprouts conjecture.
Also, we would like to modify our program and develop an application for mobile devices since the game of Sprouts is ideal for touchscreens.

 \bibliographystyle{splncs04}
 \bibliography{mybibliography}

\newpage
\appendix

\section{Graphical representation}

We will now describe individual parts of the \emph{graphical representation} $gr(P)$ of a Sprouts position $P$, which is used for correct insertion of the moves, crossing detection, and for redrawing of positions.
We need to represent vertices, edges, boundaries, and regions of a Sprouts position.
In the text, we will not explicitly distinguish between edges, boundaries, regions and their graphical representations unless it is necessary. 

\begin{figure}[!htb]
\centering
\includegraphics[scale=1]{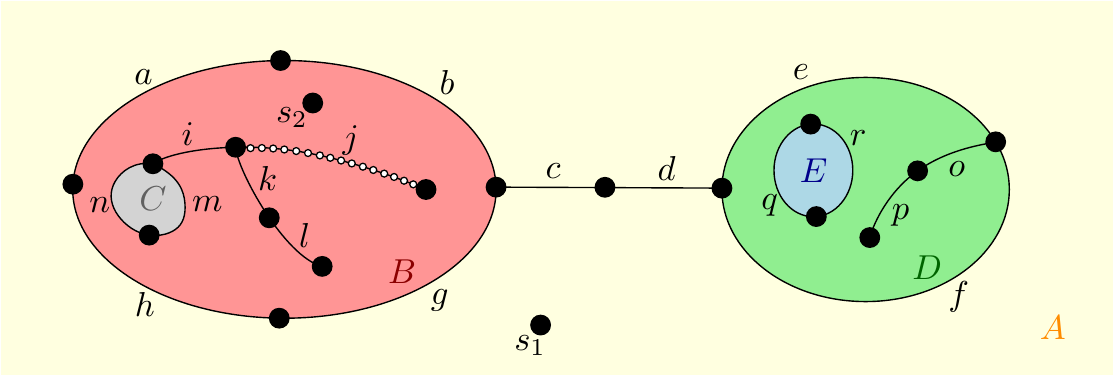}
\caption{A position with edges labelled $a$--$r$ and with regions labelled $A$--$E$.
The inner vertices of the edge $j$ are highlighted.}
\label{GraphicalRepresentationExample}
\centering
\end{figure}

We assume that the edges of $P$ are drawn as piece-wise linear arcs and that $P$ is contained in $[0,1]^2$, which represents the playing area.
An example of a position is illustrated in Figure~\ref{GraphicalRepresentationExample}.

\paragraph{Vertices} The vertices of $gr(P)$ are the endpoints of the line segments forming the piece-wise linear arcs representing the edges of $P$ or Sprouts spots that stand alone somewhere in a region of $P$.
The vertices in the latter case are called \emph{singletons}.
The \emph{game vertices} correspond to the real Sprouts spots.
They are either singletons or the endpoints of the edges of $P$.
All other vertices are called the \emph{inner vertices}.

\paragraph{Edges}
An edge $e$ of the position $P$ is represented by a sequence $gr(e) = (S,i_1, \dots, i_n, \allowbreak E)$ of vertices, where the \emph{starting vertex} $S$ and the \emph{ending vertex} $E$ are game vertices corresponding to the endpoints of $e$ and $i_1,\dots,i_n$ are the inner vertices of $e$.
The ordering of $gr(e)$ is determined by the order in which the edge $e$ was drawn.
Any two consecutive vertices of $gr(e)$ are connected by a line segment that we call a \emph{small edge}.

\paragraph{Boundaries}
A boundary $\beta$ is represented by the \emph{edge-sequence} $e_1 \cdots e_n$ of all edges $e$ of $\beta$ in the order we meet them as we traverse $\beta$.
All the inner boundaries in a region of $P$ are oriented clockwise.
The border boundary is oriented counterclockwise.
In order for our edge-insertion algorithms to work properly, it is necessary that all inner boundaries of a region have the same orientation while the orientation of the border boundary is opposite.
Note that it does not matter which rotation of the edge-sequence we consider, for example, we could use an edge-sequence $e_2 e_3 \cdots e_n e_1$.
When we traverse $\beta$, we sometimes traverse an edge $e$ of $\beta$ in the order that is opposite to the orientation of $gr(e)$.
In such a case, we use $e^R$ to denote the reversed occurrence of $e$ in the edge sequence of $\beta$.
We sometimes simplify an edge-sequence $e_1 \cdots e_n$ to a \emph{vertex-sequence} $G_1 \cdots G_n$, where $G_i$ is the starting vertex of $gr(e_i)$ for $i = 1,\dots,n$.
Each term $G_i$ is called a \emph{vertex occurrence}.

\paragraph{Regions}
Every region $R$ of $P$ is represented by a set $gr(R)$ of the edge sequences $gr(\beta)$, where $\beta$ is a boundary of $R$.
We recall that every inner region $R$ of $P$ has a unique border boundary.
The regions are used to detect crossings and we organize them into a tree-like structure to more easily find given points.

\subsection{Move insertion}

To insert a drawn move into the graphical representation, we consider two types of moves depending on whether they connect vertices from the same boundary (\emph{single-boundary moves}) or not (\emph{double-boundary moves}).
An edge $e$ representing a drawn move is divided in the middle into two new edges $e'$ and $e''$ after adding a new game vertex.
These two edges are inserted into the graphical representation, but it suffices to describe only insertion of the original edge $e$ as our algorithms are the same for inserting $e$ as for inserting $e'$ and $e''$.
We also assume that the vertices we are connecting have enough lives.

\paragraph{Double-boundary move}
Let $e = (S, i_1, \dots , i_l, E)$ be the edge representing a drawn double-boundary move that connects boundaries $\alpha$ and $\beta$ with the edge-sequences $f_1 \cdots f_m$ and $g_1 \cdots g_n$, respectively.
The small edge $Si_1$ is connected to the vertex occurrence $S_j$ of the starting vertex of $f_j$ in the corresponding vertex-sequence.
Similarly, the small edge $i_lE$ is connected to the vertex occurrence $E_k$ of the starting vertex of $g_k$.
Then the double-boundary move between $\alpha$ and $\beta$ creates a new boundary $f_1 \cdots f_{j-1} e g_k \cdots g_n g_1 \cdots g_{k-1} e^R f_j \cdots f_m$.

\begin{figure}[!htb]
\centering
\begin{tabular}{ccccc}
    \includegraphics[scale=1]{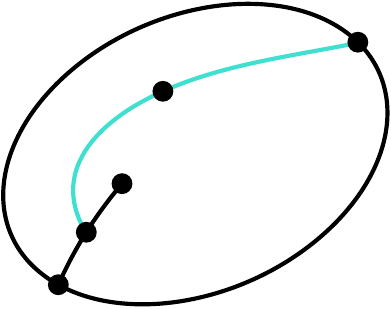} &
    &
    \includegraphics[scale=1]{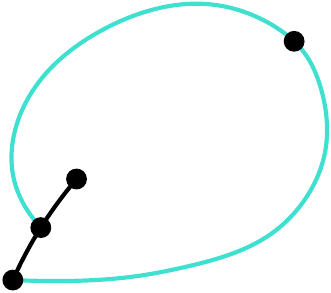} &
    &
    \includegraphics[scale=1]{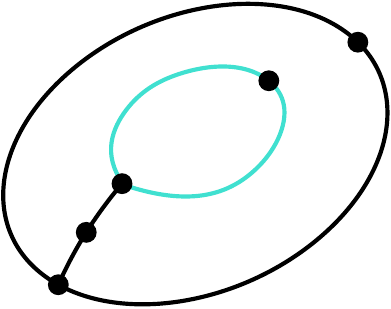} 
    \\[6pt]
(a) $e' = e^R$, $e'' = e$  && (b) $e' = e^R$, $e'' = e$ && (c) $e' = e$, $e'' = e^R$ \\[15pt]
    \includegraphics[scale=1]{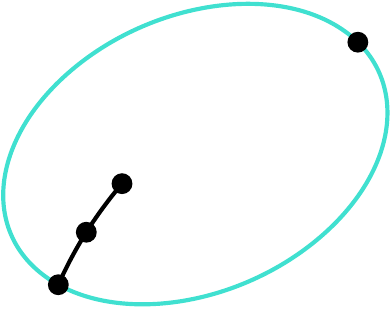} &
    &
    \includegraphics[scale=1]{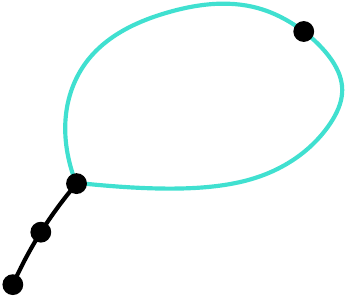} &
    &
    \includegraphics[scale=1]{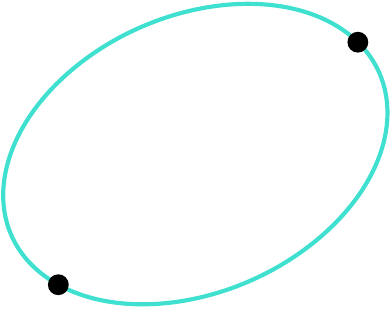} 
    \\[6pt]
    (d) $e' = e^R$, $e'' = e$  && (e) $e' = e$, $e'' = e^R$ && (f) $e' = e$, $e'' = e^R$ \\
\end{tabular}
\caption{ All types of single-boundary moves used to assign the values of $e'$ and $e''$. The moves in parts~(a)--(b) satisfy $S_j \neq E_k$ while we have $S_j = E_k$ in parts~(c)--(f).
In parts (a) and (c), the endpoints are on a border boundary and they are on an inner boundary in part~(b), (d), and (e). In part~(f), the boundary is a singleton. In part~(d), all edges of the major boundary are contained in the minor boundary which does not hold in part~(e). }
\label{SBMovesTypes}
\end{figure}

\paragraph{Single-boundary move}
Let $e = (S, i_1, \dots , i_l, E)$ be the edge representing a drawn single-boundary move that connects a vertex occurrence $S_j$ of $S$ and a vertex occurrence $E_k$ of $E$ on a boundary $\alpha$ with the edge-sequence $f_1 \cdots f_j \cdots f_k \cdots f_m$ in a region $R$.
If $S_j = E_k$, then we assume that the drawn loop has the counterclockwise orientation.
The single-boundary move on $\alpha$ creates two new boundaries $\beta$ with the edge-sequence $f_j \cdots f_{k-1} e'$, called the \emph{major boundary}, and $\gamma$ with the edge-sequence $f_1 \cdots f_{j-1} e'' f_k \cdots f_m$, called the \emph{minor boundary}, where each of $e'$ and $e''$ is equal to either $e$ or $e^R$.
The actual values of $e'$ and $e''$ are decided using a small case analysis that is shown in Figure~\ref{SBMovesTypes} where six different types of single-boundary moves are listed.

\begin{figure}[!htb]
\centering
\begin{tabular}{ccc}
\includegraphics[scale=1]{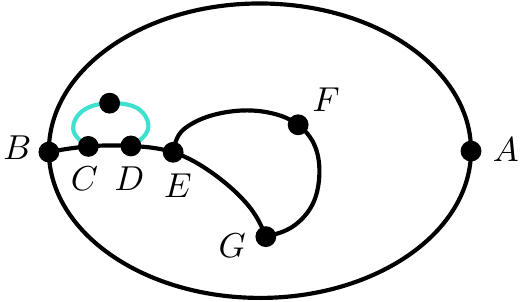} &
&
\includegraphics[scale=1]{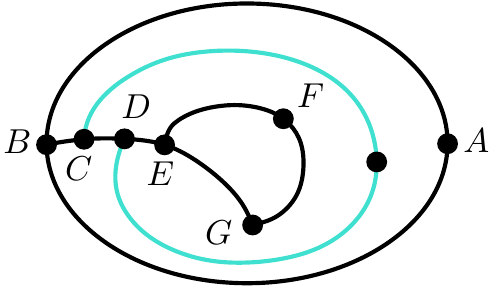} 
\\[6pt]
Connecting the first occurrence of $C$ && Connecting the first occurrence of $C$ \\
and the first occurrence of $D$. && and the second occurrence of $D$. \\ 
\end{tabular}
\caption{An example of two non-equivalent moves on a boundary with the vertex-sequence $ABCDEFGEDCB$. Note that both moves connect same pair of vertices. }
\label{ConnectedOccurrences}
\centering
\end{figure}

\paragraph{} 
To insert an edge $e$ into the graphical representation, we need to correctly identify the vertex occurrences of its endpoints in the boundary or the boundaries that $e$ is attached to.
It is important to correctly identify the occurrences as otherwise we might get a different position; see Figure~\ref{ConnectedOccurrences} for two non-equivalent moves that differ only in the occurrences of vertices they connect.
 
Consider boundaries $\alpha$ and $\beta$ and an edge $e$ connecting a game vertex $S$ of $\alpha$ and a game vertex $E$ of $\beta$. 
We need to identify the right occurrence of $S$ and $E$ in order to determine the sides from which the edge $e$ approaches $\alpha$ and $\beta$.
By symmetry, it suffices to show how to correctly identify the occurrence of $S$ in $\alpha$.
Let $e_1 \cdots e_n$ be the edge-sequence of $\alpha$ and let $Sp$ be the first small edge of $e$.
It follows from the fact that our inner boundaries have the clockwise orientation and the border boundaries have the counterclockwise orientation that a vertex occurrence $S_k$ of $S$ in $\alpha$ is correct if ${\rm deg}(S_k) \leq 1$ or if $p$ lies on the left side of $\alpha$ with the respect to the orientation of $\alpha$.

\section{Constants for the redrawing algorithm}

Since the correctly balanced values of the constants are the core of the redrawing algorithm, we list all their values that are used in our program in Table~\ref{ConstantsValues}.

\begin{table}
\begin{center}
\begin{tabular}{ |c|c|c| }
\hline
\;Symbol\; & Name & Value \\
\hhline{|=|=|=|}
$w$ & \makecell{the width of a small edge} & {$\!\begin{aligned}  
               0.023 - 0.0004 \cdot |V_G| \qquad &|V_G| \leq 40 \\
               0.007 \qquad &\text{otherwise} \end{aligned}$}\\
\hline  
$l_{opt}$ & \makecell{the optimal length of \\ a small edge} & {$\!\begin{aligned}  
               0.045 - 0.0008 \cdot |V_G| \qquad &|V_G| \leq 40 \\
               0.013 \qquad &\text{otherwise} \end{aligned}$}\\ 
\hline               
$l_{mer}$ & \makecell{the limit length for \\ merging adjacent small \\ edges together } & $l_{opt}$ \\
\hline  
$l_{sub}$ & \makecell{the limit length for \\ subdividing a small edge} & $1.8 \cdot l_{opt}$ \\
\hline
$\delta$ & \makecell{the constant for \\the attraction force } & $l_{opt}$ \\
\hline  
$\beta_{u, v}$ & \makecell{the constant for \\ the repulsion force \\ between vertices \\ $u$ and $v$ } & {$\!\begin{aligned}  
                    4 \delta \qquad &uv \in E \text{ and } uv \subseteq R_O \\
                    3 \delta \qquad &uv \in E \text{ and } uv \not \subseteq R_O\\
                    \;\;2 \delta \qquad &uv \notin E \text{ and } \{u,v\} \cap V_G \neq \emptyset\;\; \\
                    \delta \qquad &\text{otherwise}
                    \end{aligned}$} \\
\hline  
$\gamma$ & \makecell{the constant for the edge- \\ vertex repulsion force} & $10 \cdot w$ \\
\hline  
$M_{max}$ & \makecell{the maximal movement \\ in a single iteration} & $0.01$ \\
\hline  
\end{tabular}
\end{center}
\caption{The values of the constants in the redrawing algorithm. We use $V_G$ to denote the set of the game vertices of the position $P$ and $R_O$ to denote the outer region of $P$.}
\label{ConstantsValues}
\end{table}

Note that the width and the optimal length of small edges depend on the number of game vertices.
This is because a position runs out of available space with the increasing number of game vertices, or equivalently with the increasing number of moves.
Thus, we need to gradually decrease the width of small edges to saves some space.
We also shorten the length of edges since we have to make drawings finer as the position contains a higher number of vertices.

Note that $\beta_{u,v}$ can acquire four different values depending on the vertices $u$ and $v$.
In general, the repulsion force is amplified if one of the vertices is a game vertex.
To make the edges long enough, the repulsion force is strengthened even more between game vertices on the same edge.
Furthermore, to make edges in the outer region longer than in inner regions, we make these forces the strongest in this case.

\section{Drawing a computer move}

Here, we synchronize the graphical representation $gr(P)$ with the string representation $sr(P)$ of a position $P$ so that we can draw a computer move found with $sr(P)$ into $gr(P)$.
This is one of the most difficult steps we had to deal with and as far as we know, it is not fully described in the literature.
Browne~\cite{InteractiveSprouts} sketched out the idea of using \emph{Delaunay triangulations} and \emph{Voronoi diagrams}.
Although his solution works for the $n$-spot positions, there are several missing parts for more complicated positions, for example, his solution does not take into account the border boundary of a region in which the move is drawn.
So we apply our own new \emph{spindle method}.

\begin{figure}[!htb]
\centering
\includegraphics[scale=0.7]{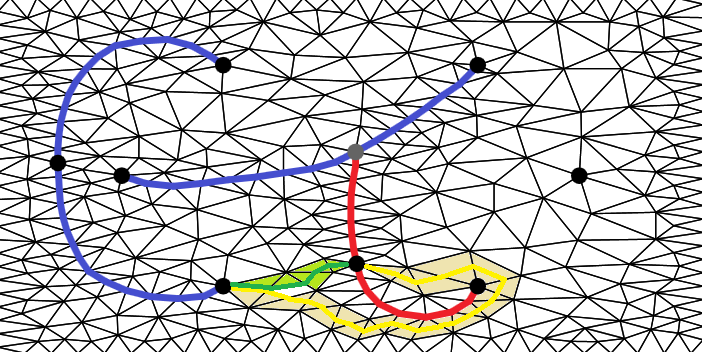}
\caption{A region triangulated by a CCDT $\mathcal{T}$ and of two paths (surrounded by shaded triangles) in $G_{\mathcal{T}}$ that connect different vertex occurrences of the same vertices.}
\label{SearchingPathExampleApen}
\end{figure}

We also use Delaunay triangulations for computer's drawing.
A \emph{constrained Delaunay triangulation} of a region of $P$ is a generalization of the Delaunay triangulation that can force the triangulation to use some edges of $G$ called the \emph{constrained edges}.
The \emph{constrained conforming Delaunay triangulation} (CCDT) of a region of $P$ is a constrained Delaunay triangulation using \emph{Steiner points} to meet given constraints on the minimum angle and the maximum area of the triangles.
We use the CCDT to triangulate a region whose edges are all constrained; see Figure~\ref{SearchingPathExampleApen}.
Let $p_1$ and $p_2$ be two non-Steiner points of a CCDT~$\mathcal{T}$.
We define a plane graph $G_\mathcal{T}=G_\mathcal{T}(p_1,p_2)$ by letting the vertices of $G_\mathcal{T}$ be the points $p_1$, $p_2$ and the midpoints of all non-constrained edges of $\mathcal{T}$.
Two vertices of $G_\mathcal{T}$ are connected by an edge in $G_\mathcal{T}$ if they lie in the same triangle of $\mathcal{T}$ but not on the same edge of $\mathcal{T}$.
If $p_1$ and $p_2$ lie in the same triangle $T$, then we connect them through the center of gravity of $T$ in $G_\mathcal{T}$.

We distinguish two types of moves depending on whether they connect vertices from different boundaries or from the same boundary of $P$.
Consider two different boundaries $A_1 \cdots A_i \cdots A_m$ and $B_1 \cdots B_j \cdots B_n$ of $P$.
The \emph{double-boundary move} that connects two \emph{vertex occurrences} $A_i$ and $B_j$ creates a new boundary $A_1 \cdots A_i Z \allowbreak B_j \cdots B_n \allowbreak B_1 \cdots B_j Z A_i \cdots A_m$ where $Z$ is the newly added vertex.
Let $A_1 \cdots A_i \cdots \allowbreak A_j \cdots A_n$ be a boundary in a region $R$ with boundaries partitioned into sets $\mathcal{B}_{major}$, $\mathcal{B}_{minor}$, and $\{\beta\}$.
The \emph{single-boundary move} that connects (not necessarily different) vertex occurrences $A_i$ and $A_j$ and separates the boundaries $\mathcal{B}_{major}$ from the boundaries of $\mathcal{B}_{minor}$ splits $R$ into the \emph{major region} $A_i \cdots A_j Z.\mathcal{B}_{major}$ and into the \emph{minor region} $A_1 \cdots A_i Z A_j \cdots A_n.\mathcal{B}_{minor}$.
We need to distinguish between $\mathcal{B}_{major}$ and $\mathcal{B}_{minor}$ since swapping them in the major and minor regions can create a non-equivalent string.

\paragraph{Searching a path inside a triangulation}
It follows from the definition of $G_\mathcal{T}$ that any two non-Steiner points $p_1, p_2$ of $\mathcal{T}$ are reachable by a path in $G_\mathcal{T}$ and no path between points $p_1$ and $p_2$ crosses any constrained edge of $\mathcal{T}$.
To find the shortest path between $p_1$ and $p_2$, we simply use the \emph{breadth-first search} (BFS).
Our path-searching algorithm also distinguishes different vertex occurrences of the same vertex.
This is important, as connecting wrong vertex occurrences may produce different positions; see Figure~\ref{SearchingPathExampleApen}.
We often use a modification of the BFS that searches the shortest path between two not necessarily disjoint sets of non-Steiner points of a triangulation $\mathcal{T}$.

\paragraph{Drawing a double-boundary move}
To draw a double-boundary move $m$ between two vertex occurrences $A_i$ and $B_j$ in a region $R$, we construct a triangulation $\mathcal{T}$ of $R$ and we let $m$ be the shortest path in $G_\mathcal{T}(A_i,B_j)$ between $A_i$ and $B_j$.

\paragraph{Drawing a single-boundary move}
Consider a single-boundary move $m$ connecting vertex occurrences $B_i$ and $B_j$ with $i \leq j$ on a boundary $\beta$ of a region $R$ with boundaries $\mathcal{B}$ that splits the boundaries $\mathcal{B} \setminus \{\beta\}$ into a major partition $\mathcal{B}_{major}$ and a minor partition $\mathcal{B}_{minor}$. Drawing of $m$ is much more complicated since we have to correctly split $\mathcal{B}$ into $\mathcal{B}_{major}$ and $\mathcal{B}_{minor}$.
The first step is to connect all the inner boundaries from $\mathcal{B} \setminus \{\beta\}$ by a curve called \emph{spindle} that starts and ends in the border boundary (or the border of the playing area if $R$ is the outer region); see Figure~\ref{Spindle}.
Then we intertwine $m$ with the spindle so that the partitions $\mathcal{B}_{major}$ and $\mathcal{B}_{minor}$ are on the correct sides of $m$; see Figure~\ref{Enfolding}.
Intertwining $m$ also uses triangulations and requires a lot of technical steps that are described in detail below.
From now on, we use the term \emph{border boundary} also for the border of the playing area.

\begin{figure}[!htb]
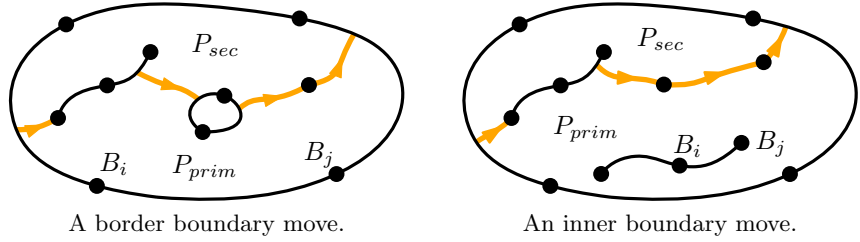

\centering
\begin{tabular}{ccc}
\includegraphics[scale=1]{border_spindle.pdf} & \;\;\;\;\;\;
    &
    \includegraphics[scale=1]{inner_spindle.pdf} \\
    A border boundary move. && An inner boundary move. \\
\end{tabular}
\caption{Spindles (heavier orange curves) that split the region into $P_{prim}$ and~$P_{sec}$.}
\label{SpindleApen}
\end{figure}

\paragraph{Setting up the spindle}
The spindle starts at an arbitrary vertex occurrence of the border boundary and leads to the closest vertex of an inner boundary from $\mathcal{B} \setminus \{\beta\}$.
Then it continues from a vertex of the last visited boundary to the closest vertex of a non-visited inner boundary from $\mathcal{B} \setminus \{\beta\}$ until we visit all the inner boundaries from $\mathcal{B} \setminus \{\beta\}$.
We end the spindle by connecting it to the closest non-visited occurrence of the border boundary; see Figure~\ref{SpindleApen}.
If there is no first inner boundary that could be connected by the spindle, we simply lead the spindle between two different occurrences of the border boundary.
It then does not matter which occurrences are chosen since the spindle is not used in this case.
If $\beta$ is the border boundary, then the starting and the ending occurrence cannot be equal or lie between the occurrences $B_i$ and $B_j$.

Each part of the spindle is constructed using a triangulation $\mathcal{T}$ of the gradually modified region $R$ and the path-searching algorithm over the graph $G_\mathcal{T}$.
The spindle divides the border polygon of $R$ into the \emph{primary polygon} $P_{prim}$, which is the polygon that contains $\beta$ or the occurrences $B_i$ and $B_j$ if $\beta$ is the border boundary, and the \emph{secondary polygon} $P_{sec}$.
The orientation of the spindle is opposite to the orientation induced by the counterclockwise orientation of $P_{prim}$.

\paragraph{Intertwining the spindle}
Let $\mathcal{C}$ be a set of some of the inner boundaries visited by the spindle $s$. 
We intertwine the move $m$ with $s$ using an \emph{enfolding} of $\mathcal{C}$; see Figure~\ref{Enfolding}.

To enfold $\mathcal{C}$, we first draw a curve in $P_{prim}$ from $v_0 = B_i$ to a vertex $v_1$ of the first segment of $s$ that precedes a boundary from $\mathcal{C}$ with respect to the orientation of~$s$.
In $P_{sec}$, we then connect $v_1$ with a vertex $v_2$ of the first segment of $s$ that precedes a boundary not in $\mathcal{C}$.
We continue connecting $v_i$ with $v_{i+1}$ like this alternatingly in $P_{prim}$ and $P_{sec}$ until we get at the end of $s$.
In this moment, if we should continue drawing in $P_{sec}$, we just draw a curve to the last segment of $s$ to get back to $P_{prim}$.
Finally, we draw a curve in $P_{prim}$ from the last connected vertex to $B_j$.
If $\mathcal{C}$ is empty, we connect $B_i$ with the first vertex of the last segment of $s$ in $P_{prim}$, then we go to the last vertex of the last segment of $s$ in $P_{sec}$, and then we return to $B_j$ in $P_{prim}$.
In a \emph{reversed enfolding} of $\mathcal{C}$, we intertwine $s$ in the opposite direction.
The boundaries from $\mathcal{C}$ are called the \emph{enfolded boundaries}.

\begin{figure}[!htb]
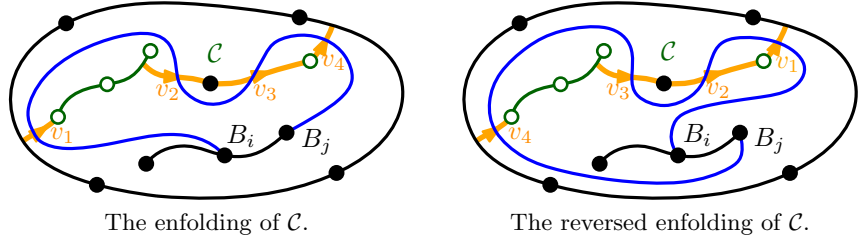

\centering
\begin{tabular}{ccc}
\includegraphics[scale=1]{enfolding.pdf} & \;\;\;\;\;\;
    &
    \includegraphics[scale=1]{reverse_enfolding.pdf} \\
    The enfolding of $\mathcal{C}$. && The reversed enfolding of $\mathcal{C}$. \\
\end{tabular}
\caption{The enfoldings of the boundaries from $\mathcal{C}$ (green curves with empty discs) if $B_i \neq B_j$. }
\label{EnfoldingApen}
\end{figure}

\begin{figure}[!htb]
\centering
\begin{tabular}{ccc}
\includegraphics[scale=1]{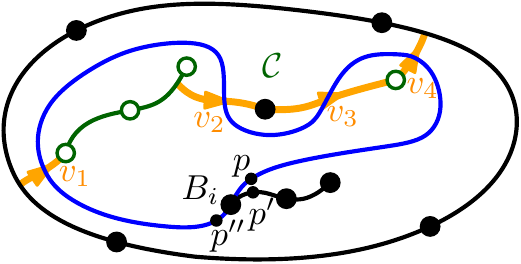} &
    &
    \includegraphics[scale=1]{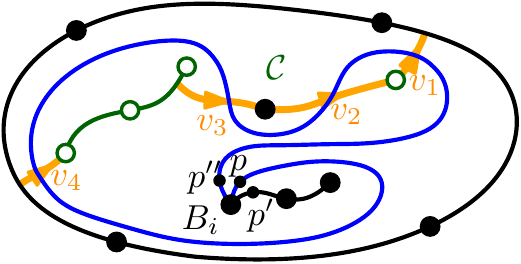} \\[6pt]
    The enfolding of $\mathcal{C}$. && The reversed enfolding of $\mathcal{C}$. \\[10pt]
\end{tabular}
\caption{The enfoldings of the boundaries from $\mathcal{C}$ (green) if $B_i = B_j$. }
\label{EnfoldingConnectionsApen}
\end{figure}

Note that if $B_i \neq B_j$, then the shape of the last curve $c$ of the spindle does affect the properties of the move.
However, this does not necessarily hold if $B_i = B_j$ since the last curve can be connected to $B_i$ in two non-equivalent ways.
In this case, we insist on $c$ being connected to $B_i$ so that the last drawn point $p$ of $c$ lies on the right side of the triple $(p', B_i, p'')$ where $p'$ is the neighboring point of $B_i$ on $\beta$ and $p''$ is the first drawn point of the first curve; see Figure~\ref{EnfoldingConnectionsApen}.

\paragraph{Choosing the right enfolding}
We have four options how to enfold the partitions $\mathcal{B}_{major}$ and $\mathcal{B}_{minor}$.
We can choose $\mathcal{C}$ as the set of inner boundaries from $\mathcal{B}_{major}$ or from $\mathcal{B}_{minor}$.
We can also apply either the enfolding or the reversed enfolding.
It follows from the definition of enfoldings that the enfolded boundaries always lie in the major region of the move $m$ whereas the reversely enfolded boundaries always lie in the minor region.
Therefore, we are left with only two options: we either enfold the inner boundaries from $\mathcal{B}_{major}$ or we reverse enfold $\mathcal{B}_{minor}$.

\begin{figure}[!htb]
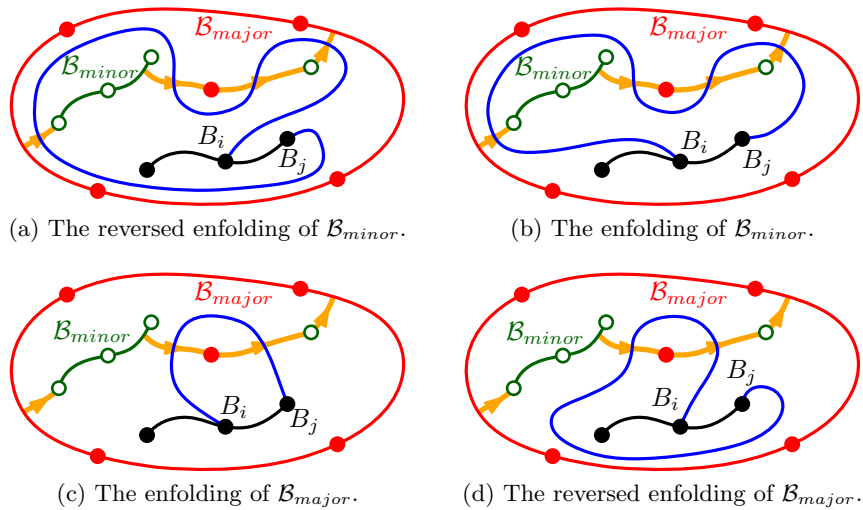

\centering
\begin{tabular}{ccc}
    \includegraphics[scale=1]{correct_enfolding.pdf} & \;\;\;\;\;\;
    &
    \includegraphics[scale=1]{incorrect_enfolding_a.pdf}\\
    (a) The reversed enfolding of $\mathcal{B}_{minor}$. & \;\;\;\;\;\;
    & (b) The enfolding of $\mathcal{B}_{minor}$. \\[10pt]
    \includegraphics[scale=1]{incorrect_enfolding_b.pdf}  & \;\;\;\;\;\;
    & \includegraphics[scale=1]{incorrect_enfolding_c.pdf}
    \\
     (c) The enfolding of $\mathcal{B}_{major}$. & \;\;\;\;\;\;
    & (d) The reversed enfolding of $\mathcal{B}_{major}$.\\
\end{tabular}
\caption{The (a) correctly and (b--d) incorrectly chosen enfoldings for drawing a move (blue) from $B_i$ to $B_j$ with the partitions $\mathcal{B}_{major}$ (full red) and $\mathcal{B}_{minor}$ (empty green). }
\label{IncorrectEnfoldingApen}
\end{figure}

Recall that $\mathcal{B}_{major}$ and $\mathcal{B}_{minor}$ can contain the border boundary which cannot be enfolded nor reversely enfolded.
Hence, the partition that contains the border boundary can never be enfolded or reversely enfolded since the border boundary would be separated from the inner boundaries in the same partition by $m$.
Thus, we are now left with a single option for enfolding.
We enfold $\mathcal{C} = \mathcal{B}_{major}$ if $\mathcal{B}_{major}$ does not contain the border boundary and we reverse enfold $\mathcal{C} = \mathcal{B}_{minor}$ otherwise; see Figure~\ref{IncorrectEnfoldingApen}.
We can see that each of the four possible enfoldings can lead to a different position.

\paragraph{Empty moves}

By the definition of intertwining the spindle, every move is always drawn across the last segment of the spindle, even when no boundary was enfolded.
However, these empty moves seem unnatural as they visit the spindle and return back without enfolding anything.
So we improve the intertwining in this case.

For moves on the border boundary that do not enfold any boundary, we simply find the shortest path in the triangulation $\mathcal{T}$ of $P_{prim}$ since all these moves always form an empty major region that does not contain any boundary on the spindle.
The only exception are empty loops connecting the same vertex occurrences.
For them, we find the first triangle in $\mathcal{T}$ containing the occurrence and then we pick two points inside the triangle that form the loop.

An empty move on a general inner boundary $\beta$ connecting occurrences $B_i$ and $B_j$ with $i \leq j$ is more advanced since the orientation of the move decides whether the boundaries on the spindle lie in the major or in the minor region.
We enforce the correct orientation by connecting $\beta$ to the primary polygon $P_{prim}$ using a so-called \emph{splitting curve} $c$.
After $c$ is drawn, the polygon $P_{prim}$ is modified so that there remains only one possible orientation of a move connecting the corresponding occurrences of $B_i$ and $B_j$ on $P_{prim}$.
We distinguish two cases.

\begin{figure}[!htb]
\centering
\begin{tabular}{ccc}
\includegraphics[scale=0.95]{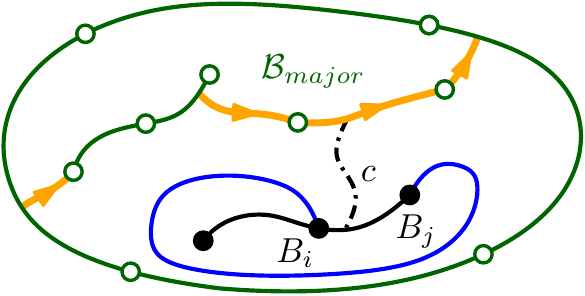} & \;\;\;\;\;\; &
    \includegraphics[scale=0.95]{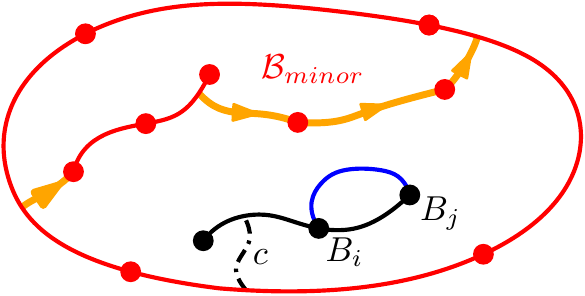}\\
    (a) $B_i \neq B_j$ with $\mathcal{B}_{major}$. & \;\;\;\;\;\; & (b) $B_i \neq B_j$ with $\mathcal{B}_{minor}$.\\[10pt]
    \includegraphics[scale=0.95]{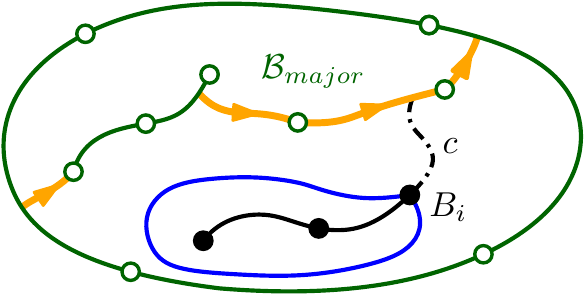} & \;\;\;\;\;\; &
    \includegraphics[scale=0.95]{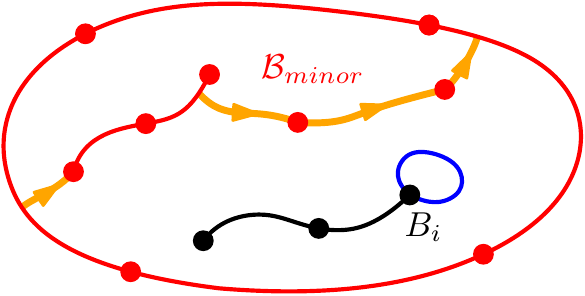}\\
    (c) $B_i = B_j$ with $\mathcal{B}_{major}$. & \;\;\;\;\;\; & (d) $B_i = B_j$ with $\mathcal{B}_{minor}$.\\
\end{tabular}
\caption{Examples of moves that do not enfold any boundaries on the spindle and that often need to use the splitting curve $c$. }
\label{EmptySBMovesApen}
\end{figure}

If $B_i \neq B_j$, we draw $c$ from the border of $P_{prim}$ to any occurrence between $B_i$ and $B_j$ if the boundaries on the spindle and the border boundary (if it exists) are from $\mathcal{B}_{major}$; see part~(a) of Figure~\ref{EmptySBMovesApen}.
Otherwise, if the boundaries are from $\mathcal{B}_{minor}$, we connect $c$ to any occurrence from the complementary range except of the occurrences $B_i$ and $B_j$ themselves; see part~(b) of Figure~\ref{EmptySBMovesApen}.
Then we simply draw a move in the modified $P_{prim}$ between $B_i$ and $B_j$.

If $B_i = B_j$, then we connect $c$ from the border of $P_{prim}$ to the occurrence $B_i$ if the boundaries on the spindle and the border boundary (if it exists) are from $\mathcal{B}_{major}$; see part~(c) of Figure~\ref{EmptySBMovesApen}.
Then we make a move from and to the vertex after the occurrence $B_i$ so that the starting occurrence on $P_{prim}$ differs from the ending occurrence on $P_{prim}$.
If the boundaries are from $\mathcal{B}_{minor}$, we simply make an empty loop using the earlier approach; see part~(d) of Figure~\ref{EmptySBMovesApen}.

\paragraph{Optimizing moves}
We use various techniques to make the moves nicer, as they should resemble moves drawn by a human player.
For example, we do not always have only a single option for enfolding.
It can happen that the border boundary is missing because it is either dead or the surrounding region is the outer region.
In this situation, we can freely choose between the enfolding of the major partition and the reversed enfolding of the minor partition.
One of the resulting two moves can be shorter than the other one. 
Since shorter moves usually seem more natural, we choose the enfolding that produces the shorter move.

We also use the fact that singletons are mutually interchangeable.
Therefore, if $n$ singletons should be enfolded, we are allowed to choose an arbitrary set of the singletons on the spindle that will be enfolded.
We enfold the $n$ singletons that are the closest to the to the point between the connected occurrences.

The drawn moves are often too close to other edges.
To solve this and to make the drawn moves smoother, we apply the redrawing algorithm just before the drawn edge is animated.
Here, we use a modification of the redrawing algorithm that moves only the vertices of the drawn edge.

\end{document}